\documentclass[a4paper,12pt]{article}

\usepackage{graphics, epsfig}

\begin{document}

\author{C. Barrab\`es\thanks{E-mail : barrabes@celfi.phys.univ-tours.fr}\\     
\small Laboratoire de Math\'ematiques et Physique Th\'eorique\\
\small  CNRS/UPRES-A 6083, Universit\'e F. Rabelais, 37200 TOURS, France\\
P.A. Hogan\thanks{E-mail : phogan@ollamh.ucd.ie}\\
\small Mathematical Physics Department\\
\small  National University of Ireland Dublin, Belfield, Dublin 4, Ireland}

\title{Detection of Impulsive Light--Like Signals in General Relativity}
\date{}
\maketitle

\begin{abstract}
The principal purpose of this paper is to study the 
effect of an impulsive light--like signal on neighbouring 
test particles. Such a signal can in general be unambiguously 
decomposed into a light--like shell of null matter and 
an impulsive gravitational wave. Our results are: (a) If there is 
anisotropic stress in the 
light--like shell then test particles initially moving in the signal front 
are displaced out of this 2-surface after encountering the signal; 
(b) For a light--like shell with no anisotropic stress 
accompanying a gravitational wave the 
effect of the signal on test particles moving in the signal front is to 
displace them relative to each other with the usual distortion due 
to the gravitational wave diminished by the presence of the light--
like shell. An explicit example for a plane--fronted signal is 
worked out. 

\end{abstract}
\thispagestyle{empty}
\newpage

\section{Introduction}\indent
The direct detection of gravitational waves will be 
realised in the near future. The detectors 
which are presently under construction are designed 
to observe signals in the frequency range from 
10 Hz to 10 kHz. The main source 
of such signals is considered to be an inspiralling binary 
neutron star system. However a large variety of 
other sources of gravitational waves exists \cite{BW}
and among them are cataclysmic events such as 
supernovae, which are expected to produce bursts of 
gravitational radiation and of null matter.

In this paper we consider impulsive gravitational 
waves and light--like shells of null matter. These 
phenomena are both impulsive light--like signals 
and mathematically correspond to a space--time geometry 
having a Riemann tensor containing a singular part 
proportional to a Dirac delta function whose support 
is a light--like hypersurface \cite{BI}. This light--like signal 
can be decomposed unambiguously \cite{BBH} into a 
gravitational wave part and a shell or matter part.

To detect such impulsive light--like signals we need to 
know their effect on the relative motion of 
neighbouring test particles. This 
is the main purpose of the present work. We consider 
a congruence of time--like geodesics in a vacuum 
space--time crossing a singular null hypersurface, 
which represents the history of the impulsive gravitational 
wave and the shell of null matter. We study the 
relative motion of the test particles 
as their world lines cross the null hypersurface.

In section 2 we present a summary of the properties 
of a singular null hypersurface aswell as the 
decomposition of the signal into an impulsive 
gravitational wave and a null shell. The details 
can be found in references \cite{BI} and \cite{BBH}. 
In section 3 the principle of detection is described and 
our main results are obtained. These are: (a) If there is 
anisotropic stress in the 
light--like shell then test particles initially moving in the signal front 
are displaced out of this 2-surface after encountering the signal; 
(b) For a light--like shell with no anisotropic stress 
accompanying a gravitational wave the 
effect of the signal on test particles moving in the signal front is to 
displace them relative to each other with the usual distortion due 
to the gravitational wave diminished by the presence of the light--
like shell. An example of a plane--fronted wave and null shell is 
worked out in detail in section 4. 

The current design of gravity wave detectors is aimed at observing 
gravitational waves that establish oscillatory motion in the 
test particles (such as the waves from an inspiralling neutron 
star system). The light--like signals we are studying displace 
the test particles but do not establish oscillatory motion. 
A modification of the detectors would be necessary to observe 
these signals.

The behavior of geodesics in space--times of impulsive gravitational 
waves has been studied in \cite{BA}--\cite{K}. The present work 
differs from these because we include a light--like shell 
in the signal and we use a local coordinate system in which the 
metric tensor is continuous across the history of the signal 
in space--time whereas in \cite{BA}--\cite{K} a distributional 
metric is used.

\setcounter{equation}{0}
\section{Fundamental Assumptions and Equations}\indent
The history of a light--like signal in a space--time 
$M$ is a null hypersurface ${\cal N}$ with the following 
properties: In a local coordinate system $\{x^{\mu}\}$ 
covering both sides of ${\cal N}$ let $(g_{\mu\nu})$ 
be the components of the metric tensor of $M$ and let 
the equation of ${\cal N}$ be $u(x^\mu )=0$. The 
normal to ${\cal N}$ has covariant components $n_\mu 
=\alpha ^{-1}u_{,\mu}$, with $\alpha$ any function of 
$x^\mu$ and the comma denoting partial differentiation 
with respect to $x^\mu$. Since ${\cal N}$ is null we have
\begin{equation} \label{2.1}
g^{\mu\nu}\,n_\mu\,n_\nu =0\ ,
\end{equation}
on ${\cal N}$. We define a transversal with contravariant 
components $N^\mu$ on ${\cal N}$ as any vector on ${\cal N}$ 
which is not tangent to ${\cal N}$. Thus
\begin{equation} \label{2.2}
N^\mu\,n_\mu =\eta ^{-1}\neq 0\  ,
\end{equation}
for some function $\eta (x^\mu)$. For ${\cal N}$ to be a 
{\it singular} null hypersurface we take the first derivatives 
of the metric tensor to jump across ${\cal N}$ and we 
furthermore assume that the jumps in these first derivatives 
take the form \cite{BI}
\begin{equation} \label{2.3}
\left [g_{\mu\nu ,\alpha}\right ]=\eta\,\gamma _{\mu\nu}\,
n_{\alpha}\ ,
\end{equation}
for some tensor $\gamma _{\mu\nu}=\gamma _{\nu\mu}$ defined 
on ${\cal N}$. We see from (\ref{2.2}) and (\ref{2.3}) 
that 
\begin{equation} \label{2.4}
N^\alpha\,\left [g_{\mu\nu ,\alpha}\right ]=\gamma _{\mu\nu}
\ .
\end{equation}
As a consequence of (\ref{2.3}) there is a jump in the 
Christoffel symbols across ${\cal N}$ given by
\begin{equation}\label{2.5}
\left [\Gamma ^\mu _{\alpha\beta}\right ]=\frac{\eta}{2}\,\left 
(\gamma ^\mu _\alpha\,n_\beta +\gamma ^\mu _\beta\,n_\alpha 
-n^\mu\,\gamma _{\alpha\beta}\right )\ .
\end{equation}
It now follows that in general there is a surface stress--
energy tensor concentrated on ${\cal N}$. This is given 
by the coefficient of $\alpha\,\delta (u)$ (where $\delta (u)$ is 
the Dirac delta function which is singular on $u=0$) in 
the expression for the Einstein tensor of $M$. The components 
$S^{\alpha\beta}$ of this stress--energy tensor are given by \cite{BI}
\begin{equation} \label{2.6}
16\pi\eta ^{-1}S^{\alpha\beta}=2\,\gamma ^{(\alpha}\,
n^{\beta )}-\gamma ^\dagger g^{\alpha\beta}-\gamma\,n^\alpha\,n^\beta\ ,
\end{equation}
with
\begin{equation} \label{2.7}
\gamma ^\alpha =\gamma ^\alpha _\beta\,n^\beta\ ,\ 
\gamma ^\dagger =\gamma ^\alpha\,n_\alpha\ ,\ \gamma 
=\gamma _{\alpha\beta}\,g^{\alpha\beta}\ .
\end{equation}
If $\{e_a\}$, with $a=1, 2, 3$, are three linearly independent 
vector fields tangential to ${\cal N}$ then on ${\cal N}$ 
the induced metric tensor has components
\begin{equation} \label{2.8}
g_{ab}=g_{\mu\nu}\,e^\mu _a\,e^\nu _b\ ,
\end{equation}
and this is, of course, degenerate. We also have
\begin{equation}\label{2.9}
\gamma _{ab}=\gamma _{\mu\nu}\,e^\mu _a\,e^\nu _b\ .
\end{equation}
It can be shown \cite{BI} that $\gamma _{ab}$ can also 
be calculated by a cut and paste technique in which the 
space--time $M$ is subdivided into $M^+(u>0)$ to the 
future of ${\cal N}$ and $M^-(u<0)$ to the past of ${\cal N}$, 
described in two different coordinate systems, and then 
re--attached on ${\cal N}$ with matching conditions which 
preserve the induced metric tensor. An example of this 
appears in section 4 below.

We can express the surface stress--energy tensor components 
in an intrinsic form as \cite{BI}
\begin{equation}\label{2.10}
16\pi\eta ^{-1}S^{ab}=(g_*^{ac}\,n^b\,n^d+g_*^{bc}\,n^a\,n^d)
\,\gamma _{cd}-\gamma ^{\dagger}\,g_*^{ab}-n^a\,n^b\,g_*^{cd}\,
\gamma _{cd}\ .
\end{equation}
Here $g_*^{ab}$ is a ``pseudo"--inverse of $g_{ab}$ in 
(\ref{2.8}) above, defined (non--uniquely) by
\begin{equation} \label{2.11}
g_*^{ab}\,g_{bc}=\delta ^a_c -\eta\,n^a\,N_c\ ,
\end{equation}
and $N_a=N_\mu \,e^\mu _a$, while $n^a$ are the coefficients 
in the expansion of the normal on the basis $\{e_a\}$:
\begin{equation}\label{2.12}
n^\mu =n^a\,e^\mu _a\ .
\end{equation}
We note from (\ref{2.8}) and (\ref{2.12}) that since $\{e_a\}$ 
are tangential to ${\cal N}$ we have $g_{ab}\,n^b=0$. 
The components $S^{\alpha\beta}$ of the surface stress--energy tensor 
in (\ref{2.6}) are recovered from the intrinsic components (\ref{2.10}) 
using
\begin{equation}\label{2.13}
S^{\alpha\beta}=S^{ab}\,e^\alpha _a\,e^\beta _b\ .
\end{equation}
If the surface stress--energy tensor (\ref{2.10}) is written 
in the form \cite{BI}
\begin{equation}\label{2.131}
\eta ^{-1}S^{ab}=\sigma\,n^a\,n^b+P\,g_*^{ab}+\Pi ^{ab}\ ,
\end{equation}
with
\begin{eqnarray}\label{2.132}
16\,\pi\,\sigma & = & -g_*^{cd}\,\gamma _{cd}\ ,\\
16\,\pi\,P & = & -\gamma ^\dagger\ ,\\
16\,\pi\,\Pi^{ab} & = & (g_*^{ac}\,n^b\,n^d+g_*^{bc}\,n^a\,n^d)
\,\gamma _{cd}\ ,
\end{eqnarray}
then we can identify $\sigma , P, \Pi^{ab}$ with the relative matter 
density, isotropic pressure and 
anisotropic stress respectively of the light--like shell. 
 
As a consequence of our basic assumption (\ref{2.3}) the Weyl tensor 
of $M$ has a term proportional to the Dirac delta function $\delta (u)$. 
The coefficient of $\delta (u)$ can be written as a sum \cite{BBH}, 
$W_{\kappa\lambda\mu\nu}+M_{\kappa\lambda\mu\nu}$ where $W_{\kappa\lambda\mu\nu}$ 
represents the gravitational wave part of the light--like signal 
and $M_{\kappa\lambda\mu\nu}$ is constructed solely from the surface 
stress--energy tensor of the signal. We note that the vectors 
$\{N^\mu\ , e^\mu _a\}$ constitute a basis for the tangent space to 
$M$ at each point of ${\cal N}$. On this basis we find that the 
components $W_{\kappa\lambda\mu\nu}\,e^\kappa _a\,e^\lambda _b\,
e^\mu _c\,e^\nu _d$ and $W_{\kappa\lambda\mu\nu}\,e^\kappa _a\,e^\lambda _b\,e^\mu _c\,
N^\nu$ of $W_{\kappa\lambda\mu\nu}$ vanish identically while 
\begin{equation}\label{2.14}
W_{\kappa\lambda\mu\nu}\,e^\kappa _a\,N^\lambda \,e^\mu _b\,N^\nu =
-\frac{1}{2}\,\eta ^{-1}\hat\gamma _{ab}\ ,
\end{equation}
with
\begin{equation}\label{2.15}
\hat\gamma _{ab}=\gamma _{ab}-\frac{1}{2}\,g_*^{cd}\,\gamma _{cd}\,
g_{ab}-2\,\eta\,n^d\,\gamma _{d(a}\,N_{b)}+\eta ^2\gamma ^\dagger\,
N_a\,N_b-\frac{1}{2}\eta ^2\gamma ^\dagger\,\left (N_\mu\,N^\mu\right )\,
g_{ab}\ .
\end{equation}
This part of $\gamma _{ab}$ satisfies $\hat\gamma _{ab}\,n^b=0=g_*^{ab}\,
\hat\gamma _{ab}$ and thus (i) it does not contribute to the 
surface stress--energy tensor (\ref{2.10}) and (ii) it has two independent 
components. We find that
\begin{equation} \label{2.16}
W_{\kappa\lambda\mu\nu}\,n^\kappa =n^a\,W_{\kappa\lambda\mu\nu}\,e^\kappa _a=0\ .
\end{equation}
Hence $W_{\kappa\lambda\mu\nu}$ is type N in the Petrov classification, 
with $n^\mu$ as four-fold degenerate principal null direction. The 
two independent components of $W_{\kappa\lambda\mu\nu}$ represent 
the two degrees of freedom of polarisation present in this gravitational 
wave part of the signal. The components $M_{\kappa\lambda\mu\nu}$ 
expressed on the basis $\{N^\mu\ , e^\mu _a\}$ are given in terms of the 
surface stress--energy tensor in \cite{BBH}. They are in general 
Petrov type II and may specialise to type III. The $M_{\kappa\lambda\mu\nu}$ 
vanish if the surface stress--energy is isotropic (i.e. if $S^{\alpha\beta}$ 
only has the final term in (\ref{2.6}) non--zero, or equivalently, 
if $S^{ab}$ only has the final term in (\ref{2.10}) non--zero) or 
if the surface stress--energy vanishes. For future reference we 
note that we have a decomposition of $\gamma _{ab}$ which, following 
(\ref{2.10}) and (\ref{2.15}), can be written as \cite{BBH}
\begin{equation}\label{2.17}
\gamma _{ab}=\hat\gamma _{ab}+\bar\gamma _{ab}\ ,
\end{equation}
with
\begin{equation}\label{2.18}
\bar\gamma _{ab}=16\pi\,\eta\,\left\{g_{ac}\,S^{cd}\,N_d\,N_b+
g_{bc}\,S^{cd}\,N_d\,N_a -\frac{1}{2}\,g_{cd}\,S^{cd}\,N_a\,N_b 
-\frac{1}{2}\,g_{ab}\,S^{cd}\,N_c\,N_d\right\}\ .
\end{equation}

\vfill\eject
\setcounter{equation}{0}
\section{Principles of Detection}\indent 
The principles upon which the detection of a light--like signal 
is carried out are based on the interaction of neighbouring test 
particles with the signal described by the geodesic deviation 
equation. In the space--time $M$ (which we shall consider to be a vacuum 
space--time except possibly on the null hypersurface ${\cal N}$) 
we consider a one--parameter family of integral curves of a vector 
field with components $T^\mu$ forming a 2--space $M_2$. We take 
$T^\mu$ to be a unit time--like vector field, so that
\begin{equation}\label{3.1}
g_{\mu\nu}\,T^\mu\,T^\nu =-1\ ,
\end{equation}
and we take the integral curves of $T^\mu$ to be time--like 
geodesics with arc length as parameter along them. Thus 
\begin{equation}\label{3.2}
\dot T^\mu \equiv T^\mu {}_{|\nu}\,T^\nu =0\ ,
\end{equation}
with the stroke denoting covariant differentiation with respect to 
the Levi--Civita connection associated with the metric tensor 
$g_{\mu\nu}$. The dot will denote covariant differentiation in 
the direction of $T^\mu$ of any tensor defined along the 
integral curves of $T^\mu$. Let $X^\mu$ be an orthogonal 
connecting vector joining neighbouring integral curves of 
$T^\mu$ and tangent to $M_2$. Thus $g_{\mu\nu}\,T^\mu\,X^\nu =0$ 
and 
\begin{equation}\label{3.3}
\dot X^\mu =T^\mu {}_{|\nu}\,X^\nu\ .
\end{equation}
It follows that $X^\mu$ satisfies the geodesic deviation equation \cite{S}
\begin{equation}\label{3.4}
\ddot{X^\mu}=-R^\mu {}_{\lambda\sigma\rho}\,T^\lambda\,X^\sigma\,T^\rho\ ,
\end{equation}
where $R^\mu {}_{\lambda\sigma\rho}$ are the components of the 
Riemann tensor of the space--time $M$. For consistency with the 
assumed jumps (\ref{2.3}) of the partial derivatives of the 
metric tensor components across ${\cal N}$, which lead to a 
possible Dirac delta function in the Riemann tensor of $M$ 
(which is singular on ${\cal N}$) we find that we can 
assume that the partial derivatives of $T^\mu$ and $X^\mu$ jump 
across ${\cal N}$ and that these jumps have the forms 
\begin{equation} \label{3.5}
\left [T^\mu {}_{,\lambda}\right ]=\eta\,P^\mu\,n_\lambda\ ,
\qquad \left [X^\mu {}_{,\lambda}\right ]=\eta\,W^\mu\,n_\lambda\ ,
\end{equation}
for some vectors $P^\mu\ , W^\mu$ defined on ${\cal N}$ (but 
not necessarily tangential to ${\cal N}$). Let $\left\{E_a\right\}$ 
be three vector fields defined along the time--like geodesics 
tangent to $T^\mu$ by parallel transporting $\{e_a\}$ 
along these geodesics. Thus 
\begin{equation}\label{3.6}
\dot E^\mu _a=0\ ,
\end{equation}
and on ${\cal N}$ we take $E_a=e_a$. The jump in the partial 
derivatives of $E^\mu _a$ must take the form
\begin{equation}\label{3.7}
\left [E^\mu _{a,\lambda}\right ]=\eta\,F^\mu _a\,n_\lambda\ .
\end{equation}
for some $F^\mu _a$ defined on ${\cal N}$.

The behavior of the orthogonal connecting vector $X^\mu$ as it 
crosses ${\cal N}(u=0)$ from the past $(u<0)$ to the future 
$(u>0)$ is important. Let $X^\mu _{(0)}$ 
denote $X^\mu$ evaluated on ${\cal N}$ (i.e. where the 
time--like geodesic, along which $X^\mu$ moves, intersects 
${\cal N}$). Let $T^\mu _{(0)}$ denote $T^\mu$ evaluated 
on ${\cal N}$. It is convenient to write
\begin{equation}\label{3.7}
X^\mu _{(0)} =X_{(0)}\,T^\mu _{(0)} +X^a _{(0)}\,e^\mu _a\ ,
\end{equation}
for some functions $X_{(0)}\ ,X^a _{(0)}$ evaluated on ${\cal N}$. We obtain the following information 
on the vectors $P^\mu\ , W^\mu\ , F^\mu _a$ appearing in 
(\ref{3.5}) and (\ref{3.7}): From (\ref{3.1}) we find that
\begin{equation}\label{3.8}
\gamma _{\mu\nu}\,T^\mu _{(0)}\,T^\nu _{(0)} +
2\,P_\mu\,\,T^\mu _{(0)} =0\ .
\end{equation} 
From (\ref{3.2}) we derive
\begin{equation}\label{3.9}
\gamma _{\alpha\beta}\,T^\alpha _{(0)}\,T^\beta _{(0)}\,
n^\mu=2\,n_\alpha\,T^\alpha _{(0)}\,\left\{P^\mu +
\gamma ^\mu _\beta\,T^\beta _{(0)}\right\}\ ,
\end{equation}
and this includes (\ref{3.8}) as a special case. 
The orthogonality of $X^\mu\ , T^\mu$ leads to
\begin{equation} \label{3.10}
\gamma _{\alpha\beta}\,X^\alpha _{(0)}\,T^\beta _{(0)} +
T^\alpha _{(0)}\,W_\alpha +X^\alpha _{(0)}\,P_\alpha =0\ .
\end{equation}
The propagation equation (\ref{3.3}) gives
\begin{equation}\label{3.11}
W^\mu =X_{(0)}\,P^\mu\ .
\end{equation}
If (\ref{3.11}) is substituted in (\ref{3.10}) then the resulting 
equation can be obtained from (\ref{3.9}). The expression 
for $W^\mu$ obtained using (\ref{3.9}) and (\ref{3.11}) 
can be derived directly from the geodesic deviation equation (\ref{3.4}). 
Finally (\ref{3.6}) yields 
\begin{equation}\label{3.12}
(n_\alpha\,T^\alpha _{(0)} )\,F^\mu _a=-\frac{1}{2}\,\gamma ^\mu 
_\alpha\,e^\alpha _a\,(n_\beta\,T^\beta _{(0)} )+\frac{1}{2}\,
n^\mu\,\left (\gamma _{\alpha\beta}\,e^\alpha _a\,T^\beta _{(0)}\right )
\ .
\end{equation}

As a consequence of (\ref{3.5}) we deduce, for small $u$, 
\begin{equation}\label{3.121}
X^\mu ={}^-X^\mu +\eta\,\alpha ^{-1}u\,\vartheta (u)\,W^\mu\ ,
\end{equation}
with ${}^-X^\mu$ in general dependent on $u$ and such that 
when $u=0$, ${}^-X^\mu =X^\mu _{(0)}$. Here $\vartheta (u)$ 
is the Heaviside 
step function which is equal to unity if $u>0$ and equal to 
zero if $u<0$. Similar equations to 
(\ref{3.121}) can be obtained for $g_{\mu\nu}\ ,T^\mu\ ,
E^\mu _a$ using (\ref{2.3}), (\ref{3.5}) and (\ref{3.7}). 
It is convenient to calculate $X^\mu$ on the basis $\left\{E^\mu _a\ , 
T^\mu\right\}$. Its component in the direction of $T^\mu$ is, 
of course, zero and its components in the directions $\left\{E^\mu _a\right \}$ 
are
\begin{equation}\label{3.14}
X_a=g_{\mu\nu}\,X^\mu\,E^\nu _a\ .
\end{equation}
Using (\ref{3.7})--(\ref{3.121}) we calculate (\ref{3.14}) to read, 
for small $u>0$, 
\begin{equation}\label{3.15}
X_a=\left (\tilde g_{ab}+\frac{1}{2}\,\eta\,\alpha ^{-1}u\,\gamma _{ab}\right )
\,X^b _{(0)}+u{}^-V_{(0)a}\ ,
\end{equation}
with $\gamma _{ab}$ given by (\ref{2.9}), ${}^-V_{(0)a}=
d{}^-X_a/du$ evaluated at $u=0$ and $\tilde g_{ab}$ given by
\begin{equation}\label{3.50}
\tilde g_{ab}=g_{ab} +\left (T_{(0)\mu}\,e^\mu _a\right )\,
\left (T_{(0)\nu}\,e^\nu _b\right )\ ,
\end{equation}
with $g_{ab}$ found in (\ref{2.8}). Although $g_{ab}$ is 
degenerate we note that $\tilde g_{ab}$ is non--degenerate. 
The final term in (\ref{3.15}) is present due to the relative 
motion of the test particles before encountering the signal. 
It would represent the relative displacement for small $u>0$ 
if no signal were present.

In parallel with (\ref{3.7}) we can write 
\begin{equation}\label{3.51}
X^\mu =X\,T^\mu +X^a\,E^\mu _a\ .
\end{equation}
Since $X^\mu\ ,T^\mu$ are orthogonal we have
\begin{equation}\label{3.52}
X=X^a\,E^\mu _a\,T_\mu\ .
\end{equation}
Substituting (\ref{3.51}) with (\ref{3.52}) into (\ref{3.14}) 
we obtain 
\begin{equation}\label{3.53}
X_a=\tilde G_{ab}\,X^b\ ,
\end{equation}
with
\begin{equation}\label{3.54}
\tilde G_{ab}=g_{\mu\nu}\,E^\mu _a\,E^\nu _b+E^\mu _a\,T_\mu\,E^\nu _b\,T_\nu\ .
\end{equation}
However on account of the parallel transport (\ref{3.2}) and 
(\ref{3.6}) of $T^\mu\ ,E^\mu _a$ we have $\tilde G_{ab}=\tilde g_{ab}$ 
and so we can write
\begin{equation}\label{3.55}
X_a=\tilde g_{ab}\,X^a\ ,
\end{equation}
and this equation can be inverted.

To see the separate effects of the wave and shell parts of the 
light--like signal on the relative position of the test particles 
we use in (\ref{3.15}) the decomposition of $\gamma _{ab}$ 
given by ({\ref{2.17}) and (\ref{2.18}). It is convenient to 
specialise the triad $\{e_a\}$ by choosing $e^\mu _1=n^\mu$ 
and $e_A$, with $A=2, 3$, orthogonal to $T^\mu _{(0)}$. It 
then follows from (\ref{2.12}) that $n^a=\delta ^a_1$ and that 
$g_{ab}=0$ except for $g_{AB}$ with $A, B=2,3$. We can specialise 
$N^\mu$ by taking $N^\mu =T^\mu _{(0)}$. Thus 
$e^\mu _A\,N_\mu=0$ and $N_a=0$ except 
for $N_1=\eta ^{-1}$. It follows from (\ref{2.11}) that we can have 
$g_*^{ab}=0$ except for $g_*^{AB}=g^{AB}$. We shall take the (2,3)--2-surface 
in ${\cal N}$ to be the signal front and we shall assume that 
before the signal arrives the test particles are in this 
(2,3)--2-surface. The metric of this space--like 2-surface has 
components $g_{AB}$ and since this is a Riemannian 2-surface 
we can choose coordinates $\{x^A\}$ such that $g_{AB}=p^{-2}\delta _{AB}$ 
with $p=p(x^A)$. Hence in these coordinates $\tilde g_{ab}=
{\rm diag}(\eta ^{-2}, p^{-2}, p^{-2})$.

Since $\hat\gamma _{ab}$ satisfies $\hat\gamma _{ab}\,
n^b=0=g_*^{ab}\,\hat\gamma _{ab}$ we have now in our special frame 
$\hat\gamma _{a1}=0$ for $a=1, 2, 3$ and  
\begin{displaymath} \label{3.16}
\left (\hat\gamma _{AB}\right )=
\left ( \begin{array}{ccc}
\hat\gamma _{22} & \hat\gamma _{23} \\
\hat\gamma _{23} & -\hat\gamma _{22} \\
\end{array} \right ) 
\end{displaymath}
It follows from (\ref{2.18}) that, in this special frame, $\bar\gamma _{ab}$ 
is given by
\begin{eqnarray}\label{3.17}
\bar\gamma _{11} & = & -8\pi\,\eta ^{-1}p^{-2}\left (S^{22}+S^{33}\right )\ , \\
\bar\gamma _{1B} & = & =\bar\gamma _{B1}=16\pi\,\eta ^{-1}S^{B1}\ ,\\
\bar\gamma _{AB} & = & -8\pi\,\eta ^{-1}p^{-2}S^{11}\,
\delta _{AB}\ .
\end{eqnarray}

Using (\ref{3.55}) we can now write (\ref{3.15}) as follows:
\begin{eqnarray}\label{3.56}
\eta ^{-2}\,X^1 & = & X_1=\eta ^{-2}\,X^1_{(0)}+
\frac{1}{2}\,\eta\,\alpha ^{-1}u\,\bar\gamma _{1B}\,
X^B_{(0)}+u{}^-V_{(0)1}\ , \\
p^{-2}\,X^A & = & X_A=p^{-2}\,X^A_{(0)}+\frac{1}{2}\,\eta\,\alpha ^{-1}u\,
\gamma _{AB}\,X^B_{(0)}+u{}^-V_{(0)A}\ .
\end{eqnarray}
We note that if there is a wave component to the signal it does 
not contribute to $X^1$. If initially in physical space the test particles 
are moving in the (2,3)--2-surface (the 
signal front) then $X^1_{(0)}=0={}^-V_{(0)1}$. In this case if 
$S^{B1}\neq 0$ then $X^1\neq 0$. 
Hence we can say that {\it if there is anisotropic stress in the 
light--like shell then test particles initially moving in the signal front 
are displaced out of this 2-surface after encountering the signal}. 

Suppose that there is no anisotropic stress $\left (S^{1B}=0\right )$ 
in the light--like shell. Now $X^1_{(0)}=0={}^-V_{(0)1}$ implies $X^1=0$ and 
we can we can rewrite (3.27) using (3.25) and the 
decomposition of $\gamma _{ab}$ in the form
\begin{equation}\label{3.19}
X^A=\left (1-4\pi\,\alpha ^{-1}u\,S^{11}\right )\,\left (
\delta _{AB}+\frac{1}{2}\,\eta\,\alpha ^{-1}u\,p^{2}\hat\gamma 
_{AB}\right )\,X^B_{(0)}\ ,
\end{equation}
for small $u$. The factor $\delta _{AB}+
\frac{1}{2}\,\eta\,\alpha ^{-1}u\,p^{2}\hat\gamma 
_{AB}$, with $\hat\gamma _{AB}$ given in matrix form above, 
describes the usual distortion effect of the wave part of the signal 
on the test particles in the signal front (see \cite{SCH} 
and the explicit example given at the end of section 4 below) 
while the presence of the light--like shell leads to an overall diminution 
factor $1-4\pi\,\alpha ^{-1}u\,S^{11}<1$. This latter inequality 
arises as follows: Since we have chosen $u$ to increase as 
one crosses ${\cal N}(u=0)$ from $M^-(u<0)$ to $M^+(u>0)$, 
we must have $\alpha ^{-1}\eta >0$ on ${\cal N}$. The relative 
energy density $\sigma$, given in (\ref{2.131}), is positive and 
so using (\ref{2.131}) in our special frame we obtain $\alpha ^{-1}
S^{11}=\alpha ^{-1}\eta\,\sigma >0$. Thus we can conclude that 
{\it for a light--like shell with no anisotropic stress accompanying a gravitational wave the 
effect of the signal on test particles moving in the signal front is to 
displace them relative to each other with the usual distortion due 
to the gravitational wave diminished by the presence of the light--
like shell}.

\setcounter{equation}{0}
\section{Plane Fronted Signal}\indent
A plane fronted signal which incorporates a gravitational wave and 
a light--like shell with anisotropic stress can easily be constructed 
propagating through flat space--time. In this case ${\cal N}$ is 
the history of a null hyper{\it plane}. In coordinates covering 
both sides of ${\cal N}$ the line--element of the space--time $M$ 
reads in this case
\begin{equation}\label{4.1}
ds^2=\left (dx +\frac{u_+}{F_v}\,dF_x\right )^2+
\left (dy +\frac{u_+}{F_v}\,dF_y\right )^2-2\,dv\,
\left (du-\frac{u_+}{F_v}\,dF_v\right )\ .
\end{equation}
Here $u_+=u\,\vartheta (u)$ with $\vartheta (u)$ the Heaviside 
step function as before. The equation of ${\cal N}$ is $u=0$. Also 
$F=F(x, y, v)$ with partial derivatives indicated by subscripts 
and with $F_v\neq 0$. For the signal with 
history $u=0$ in the space--time with line-element (\ref{4.1}) 
we find that the surface stress--energy tensor is given by 
(\ref{2.131}) with $\eta =-1$ and 
\begin{eqnarray}
\sigma & = & -\frac{1}{8\pi}\,\frac{F_{xx}+F_{yy}}{F_v}\ , \\
P & = & -\frac{1}{8\pi}\,\frac{F_{vv}}{F_v}\ , \\
\Pi ^{12} & = & -\frac{1}{8\pi}\,\frac{F_{xv}}{F_v}\ ,\qquad 
\Pi ^{13}=-\frac{1}{8\pi}\,\frac{F_{yv}}{F_v}\ ,
\end{eqnarray}
with all other components of $\Pi ^{ab}$ vanishing. The 
coefficient of the delta function in the Weyl tensor 
has a Petrov type N part ($W_{\kappa\lambda\mu\nu}$ above) 
given in Newman--Penrose notation by
\begin{equation} \label{4.6}
\hat\Psi _4=\frac{1}{2\,F_v}\,\left (F_{xx}-F_{yy}-2\,i\,F_{xy}\right )\ .
\end{equation}
There is also a Petrov type II part ($M_{\kappa\lambda\mu\nu}$ above) 
given in Newman--Penrose notation by
\begin{equation}\label{4.7}
\hat\Psi _2=\frac{1}{3}\,\frac{F_{vv}}{F_v}\ ,\qquad 
\hat\Psi _3=\frac{1}{\sqrt{2}\,F_v}\,\left (F_{xv}-i\,F_{yv}
\right )\ ,
\end{equation}
which can be written in terms of $P$ and $\Pi ^{ab}$. It is 
interesting to note that using (4.3)--(4.6) the line--element 
(\ref{4.1}) can be written in terms of $\sigma\ ,P\ ,\Pi^{ab}$ 
and $\hat\Psi _4$.

We mentioned following (\ref{2.9}) how the space--time 
model of a light--like signal could be constructed by 
a cut and paste technique. This can be seen in the case 
of (\ref{4.1}) as follows: The line--element (\ref{4.1}) 
can be transformed for $u>0$ into the manifestly flat 
form
\begin{equation}\label{4.8}
ds_+^2=dx_+^2+dy_+^2-2\,du_+\,dv_+\ ,
\end{equation}
by the transformation
\begin{eqnarray}
x_+ & = & x+u\,\frac{F_x}{F_v}\ , \\
y_+ & = & y+u\,\frac{F_y}{F_v}\ , \\
v_+ & = & F+\frac{1}{2}\,u\,F_v^{-1}\left (F_x^2+F_y^2\right )\ , \\
u_+ & = & \frac{u}{F_v}\ .
\end{eqnarray}
Thus (\ref{4.8}) is the line--element of $M^+(u>0)$. The 
line--element (\ref{4.1}) can be trivially transformed 
for $u<0$ into the manifestly flat form
\begin{equation}\label{4.13}
ds_-^2=dx_-^2+dy_-^2-2\,du_-\,dv_-\ ,
\end{equation}
by the transformation
\begin{equation}\label{4.14}
x_-=x\ ,\qquad y_-=y\ ,\qquad v_-=v\ ,\qquad u_-=u\ .
\end{equation}
Thus (\ref{4.13}) is the line--element of $M^-(u<0)$. We see 
from (4.9)--(4.11) and (4.14) that on $u=0$
\begin{equation}\label{4.15}
x_+=x_-\ ,\qquad y_+=y_-\ ,\qquad v_+=F(v_-\ , x_-\ , y_- )\ ,
\end{equation}
and these {\it matching conditions} leave the induced 
line--element of ${\cal N}(u=0)$ invariant:
\begin{equation}\label{4.16}
dx_+^2+dy_+^2=dx_-^2+dy_-^2\ .
\end{equation}
We interpret (\ref{4.8})--(\ref{4.16}) as follows: We have 
subdivided the space--time $M$ into two halves, $M^+(u>0)$ 
and $M^-(u<0)$, and we have re--attached the halves on ${\cal N}$ 
with the mapping (\ref{4.15}) which preserves the line--element 
induced on ${\cal N}$. This is an example of the cut and paste 
procedure mentioned in section 2.

A simple explicit example of (\ref{3.19}) is provided by specializing 
(\ref{4.1}) with the choice
\begin{equation}\label{4.17}
F(x, y, v)=v-\frac{a}{2}\,(x^2+y^2)+\frac{b}{2}\,(x^2-y^2)+c\,x\,y\ ,
\end{equation}
with $a, b, c$ constants and $a>0$. This is a homogeneous signal 
with vanishing stress and relative energy density
\begin{equation}\label{4.18}
\sigma =\frac{a}{4\,\pi}\ ,
\end{equation}
in the light--like shell and with constant amplitude of each 
component of the gravitational wave since the entries in 
\begin{displaymath} \label{4.19}
\left (\hat\gamma _{AB}\right )=
\left ( \begin{array}{ccc}
2\,b & 2\,c \\
2\,c & -2\,b \\
\end{array} \right )\ , 
\end{displaymath}
are constants. Assuming zero relative velocity for the 
test particles before the signal arrives, (\ref{3.19}) reads
\begin{eqnarray}\label{4.20}
X^2 & = & (1-a\,u)\,(1+b\,u)\,\left (X^2_{(0)}+c\,u\,X^3_{(0)}\right )\ ,\\
X^3 & = & (1-a\,u)\,(1-b\,u)\,\left (c\,u\,X^2_{(0)}+X^3_{(0)}\right )\ .
\end{eqnarray}
Hence particles at rest on the circle $\left (X^2_{(0)}\right )^2+
\left (X^3_{(0)}\right )^2={\rm constant}$, before encountering 
the light--like signal, undergo a small displacement after 
encountering the signal which is composed of: (1) a rotation 
through the small angle $c\,u$ and (2) a deformation into an 
ellipse with semi axes of lengths $(1-a\,u)\,(1+b\,u)$ and 
$(1-a\,u)\,(1-b\,u)$. If the light--like shell is not part 
of the signal (i.e. if $a=0$) then the semi axes of the 
ellipse are {\it not} shortened by the factor $(1-a\,u)$. If 
$a=c=0$ in (\ref{4.17}) then (\ref{4.1}) coincides with 
the homogeneous plane impulsive wave of Penrose \cite{P}

\vskip 2truepc
\noindent
This collaboration has been funded by the Minist\`ere des Affaires 
\'Etrang\`eres, D.C.R.I. 220/SUR/R.

\end{document}